\documentclass[prl,aps,twocolumn,showpacs,superscriptaddress]{revtex4}
\usepackage{graphicx}
\usepackage{dcolumn}

\begin{document}

\title{Collective Oscillations of Vortex Lattices in  Rotating Bose-Einstein Condensates}

\author{T. Mizushima}
\affiliation{Department of Physics, Okayama University,
Okayama 700-8530, Japan}
\author{Y. Kawaguchi}
\affiliation{Department of Physics, Graduate School of Science, Kyoto University,
Kyoto 606-8502, Japan}		
\author{K. Machida}
\affiliation{Department of Physics, Okayama University,
Okayama 700-8530, Japan}
\author{T. Ohmi}
\affiliation{Department of Physics, Graduate School of Science, Kyoto University,
Kyoto 606-8502, Japan}		
\author{T. Isoshima}
\affiliation{Materials Physics Laboratory, Helsinki University of Technology,
P.~O.~Box 2200 (Technical Physics), FIN-02015 HUT, Finland}
\author{M. M. Salomaa}
\affiliation{Materials Physics Laboratory, Helsinki University of Technology,
P.~O.~Box 2200 (Technical Physics), FIN-02015 HUT, Finland}
\date{\today}

\begin{abstract}

The complete low-energy collective-excitation spectrum of vortex lattices is discussed  
for rotating Bose-Einstein condensates (BEC) by solving the Bogoliubov-de Gennes (BdG)
equation, yielding, e.g., the Tkachenko mode recently observed at JILA. 
The totally symmetric subset of these modes includes the transverse shear, 
common longitudinal, and differential longitudinal modes.
We also solve the time-dependent Gross-Pitaevskii (TDGP) equation to simulate the actual
JILA experiment, obtaining the Tkachenko mode and identifying a pair of breathing modes. 
Combining both the BdG and TDGP approaches allows one to unambiguously 
identify every observed mode.

\end{abstract}


\maketitle

Owing to their fundamental significance for superfluidity, quantized vortices 
have attracted widespread interest in many different 
physical systems ranging from superconductors, 
superfluid $^3$He and $^4$He, to neutron stars \cite{donnelly} 
and extending to cosmology. 
Recently, vortices have been created 
in dilute alkali-atom gases using three different methods: phase imprinting \cite{nist}, 
topological phase engineering \cite{mit} and optical spoon stirring \cite{ens}. 
The former two were first predicted theoretically \cite{holland,nakahara}.  
Several groups are now able to routinely prepare
a vortex array with hundreds of vortices in a BEC.

In the mid 1960's, Tkachenko \cite{tkachenko} predicted that 
a vortex lattice would sustain a collective 
vortex-oscillation mode in which the vortex cores move elliptically
around the equilibrium positions. Subsequent theoretical
advances were made \cite{baym,fetter,campbell,donnelly,sonin}.
In superfluid $^4$He, the Tkachenko modes were first observed in 1982 \cite{andereck}.

Recently, Coddington {\it et al.} \cite{coddington} succeeded 
in observing the Tkachenko mode (TK) 
in a rotating BEC of  $^{87}$Rb. The TK wave was excited by removing condensate
from the central region of the rotating cloud. They created the lowest and second-lowest
TK modes and measured their energies $\omega_{1,0}$ and $\omega_{2,0}$ as functions 
of the rotation frequency $\Omega$. They also discovered various new phenomena,
some of which were explained by two groups \cite{anglin,baym2}:
Anglin and Crescimanno \cite{anglin} extended the previous hydrodynamic description for infinite
systems to a finite harmonically trapped system,
and Baym \cite{baym2} discovered subtle effects due to the finite compressibility.

Here we choose a different approach to this problem: We wish to treat the whole low-lying 
collective-excitation spectrum of trapped BECs; not only the TK mode but also the other important
modes and their intrinsic relationships. We construct a first-principles theory,
namely the Bogoliubov-de Gennes equation (BdG) coupled with the Gross-Pitaevskii equation (GP).
The set of equations within the Bogoliubov framework is regarded as the fully microscopic theory 
of dilute Bose gases, in the sense that there remain no adjustable parameters
once we have fixed the atomic species and the atomic number.
Thus it is quite reasonable to expect that this formalism must well apply to analyzing the
TK mode and also to provide the complete spectral features of the low-energy excitations, beyond
any limitations of hydrodynamics which has thus far been the only way to describe the TK mode.
We can then better characterize the various modes, such as the 
three classes of compressional modes: the transverse, common longitudinal 
and the differential longitudinal waves.
These are characteristic to the two-component system consisting of 
a vortex lattice and the superfluid.
Some of the selected common longitudinal modes in a vortex lattice (the breathing and
quadrupole modes) have been recently examined \cite{stringari} within hydrodynamics.
This theory may help one to gain improved understanding of related problems, 
such as vortex pinning and vortex melting in superconductors 
which have thus far only been analyzed phenomenologically.
 
In a frame rotating with the frequency $\Omega$, 
the time-dependent Gross-Pitaevskii equation  (TDGP) may be expressed 
for the condensate wavefunction $\psi$ \cite{leggett} as
\begin{eqnarray}
i \hbar \frac{\partial}{\partial t} \psi =
	\left[
				\frac{\hat{\mbox{\boldmath $p$}} ^2 (\Omega)}{2m} 
				+ V _{\rm eff} ({\bf r},\Omega)
				-\mu + g |\psi|^2
	\right] \psi, 
\label{eq:2dgp}
\end{eqnarray}
where 
$\hat{\mbox{\boldmath $p$}} (\Omega) 
= - i\hbar \nabla - m \Omega \hat{z} \times {\bf r}$
and the effective confining potential is  
$V_{\rm eff} ({\bf r},\Omega)=\frac{1}{2}m(\omega _r^2 - \Omega ^2)r^2$
with the radial trap frequency $\omega _r$.

In order to study the collective oscillations of vortex lattices microscopically,
we first consider the equation of motion for a small perturbation 
around the stationary state $\phi _g$, {\it i.e.}, 
$\psi ({\bf r}, t) = \phi _g ({\bf r}) + u_{\bf q} ({\bf r}) e^{-i\omega _{\bf q} t} 
- v^{\ast}_{\bf q} ({\bf r}) e^{i\omega _{\bf q} t}$,
where the equilibrium state $\phi _g$ is determined by the stationary GP equation.
By retaining terms up to first order in $u$ and $v$, we derive the BdG equation:
\begin{eqnarray}
	\left(
		\begin{array}{cc}
			\mathcal{L}  ({\bf r}, \Omega) & - g \phi ^2 _g ({\bf r}) \\
			g \phi^{\ast 2} _g ({\bf r}) & -\mathcal{L} ({\bf r}, - \Omega)
		\end{array}
	\right)
	\left(
		\begin{array}{c}
			u_{\bf q} ({\bf r}) \\ v_{\bf q} ({\bf r}) 
		\end{array}
	\right)
	= \hbar \omega _{\bf q}
	\left(
		\begin{array}{c}
			u_{\bf q} ({\bf r}) \\ v_{\bf q} ({\bf r}) 
		\end{array}
	\right) , 
\label{eq:bdg}
\end{eqnarray}
where
$\mathcal{L} ({\bf r}, \Omega)=  \hat{\mbox{\boldmath $p$}} ^2 (\Omega) / 2m
 + V _{\rm eff} ({\bf r},\Omega) -\mu + 2 g | \phi _g ({\bf r}) | ^2$.
We consider the JILA experiment with $2.0 \times 10^6$ atoms of $^{87}$Rb 
confined in a trap with the radial frequency $\omega _{r}/2\pi=8.3$ Hz 
and an axial one $\omega _z /2\pi= 5.2$ Hz \cite{coddington}.
Employing Thomas-Fermi (TF) theory and the assumption of solid-body rotation, 
the condensate aspect ratio is given as 
$\lambda _{\rm TF} \equiv R_{\rm TF} / Z_{\rm TF} 
\propto (\omega^2_r-\Omega^2)^{-1/2}$ 
where $R_{\rm TF}$ and $Z_{\rm TF}$ are the condensate lengths along the $r$- and $z$-axes.
This relation allows one to assume the system to constitute a two-dimensional geometry 
at high rotation frequencies. Under this assumption, we introduce the linear density $n_z(\Omega)$ 
along the $z$ axis. Therefore, the equilibrium state $\phi _g$ must fulfill the normalization condition 
$n_z(\Omega)=\int \int |\phi _g|^2 dxdy$, where
the linear density is obtained as $n_z(\Omega)=R^4_{\rm TF}/16ad^4_r$,
with $a$ an $s$-wave scattering length, 
and $d^2_r\equiv \hbar / ( m \sqrt{\omega^2 _r-\Omega^2} )$.
We discretize the two-dimensional space typically into a  
$300^2 \sim 1000^2$ mesh to solve the TDGP and BdG equations.

Here, since we consider a vortex array with sixfold symmetry, 
the wavefunction of the stationary state obeys the condition, 
$\phi _g (R^{n} {\bf r}) = \phi _g ({\bf r}) e^{i n\pi/3}$, 
where $R^{n} {\bf r}$ describes a rotation $n\pi/3$ ($n$ integer) 
around the center of the trap, {\it i.e.}, 
$R^{n} {\bf r} = (x\cos{(n\pi/3)} - y \sin{(n\pi/3)}, x\sin{(n\pi/3)}+y\cos{(n\pi/3)})$.
We then obtain the following relation from Eq.~(\ref{eq:bdg}): 
$u_{{\bf q},m} (R^{n} {\bf r}) = u_{\bf q} ({\bf r}) \exp{\left[\frac{in\pi}{3}(m+1) \right]}$ 
and $v_{{\bf q},m} (R^{n} {\bf r}) = v_{\bf q} ({\bf r}) \exp{\left[\frac{in\pi}{3}(m-1) \right]}$, 
where $m= 0, \pm 1, \pm 2, 3$.
In order to classify the collective excitations,
we introduce the following classifying function: 
$F^{(u)}_{\bf q}(m) = \int d{\bf r}
  u_{\bf q}^{\ast}({\bf r}) 
  \left[ \sum ^5 _{n = 0} u_{{\bf q}, m} (R^{n} {\bf r}) \right]
  / \int d{\bf r} | u_{\bf q} ({\bf r}) |^2 $, 
which tends to $1$ for a suitable $m$ and 0 for the others.
Furthermore, we define the average angular momentum as
$q_{\theta} 
\equiv [ \langle L_z \rangle _u + \langle L_z \rangle _v - \langle \hat{L}_z \rangle _{\phi _g}]
/ \int d{\bf r} [|u({\bf r})|^2 + |v({\bf r})|^2]$, 
where 
$\langle L_z \rangle _{\phi _g}
\equiv \int d{\bf r} \phi _g ^{\ast} \hat{L}_z \phi _g / \int d{\bf r} |\phi _g|^2$, and 
$\langle L_z \rangle _u \equiv \int d{\bf r}  u^{\ast} \hat{L}_z u$ \cite{isoshima}.
In an axisymmetric situation, $m$ and $q_{\theta}$ 
merge into the same integer quantum number.

\begin{figure}[t]
(a) \includegraphics[width=3.7cm]{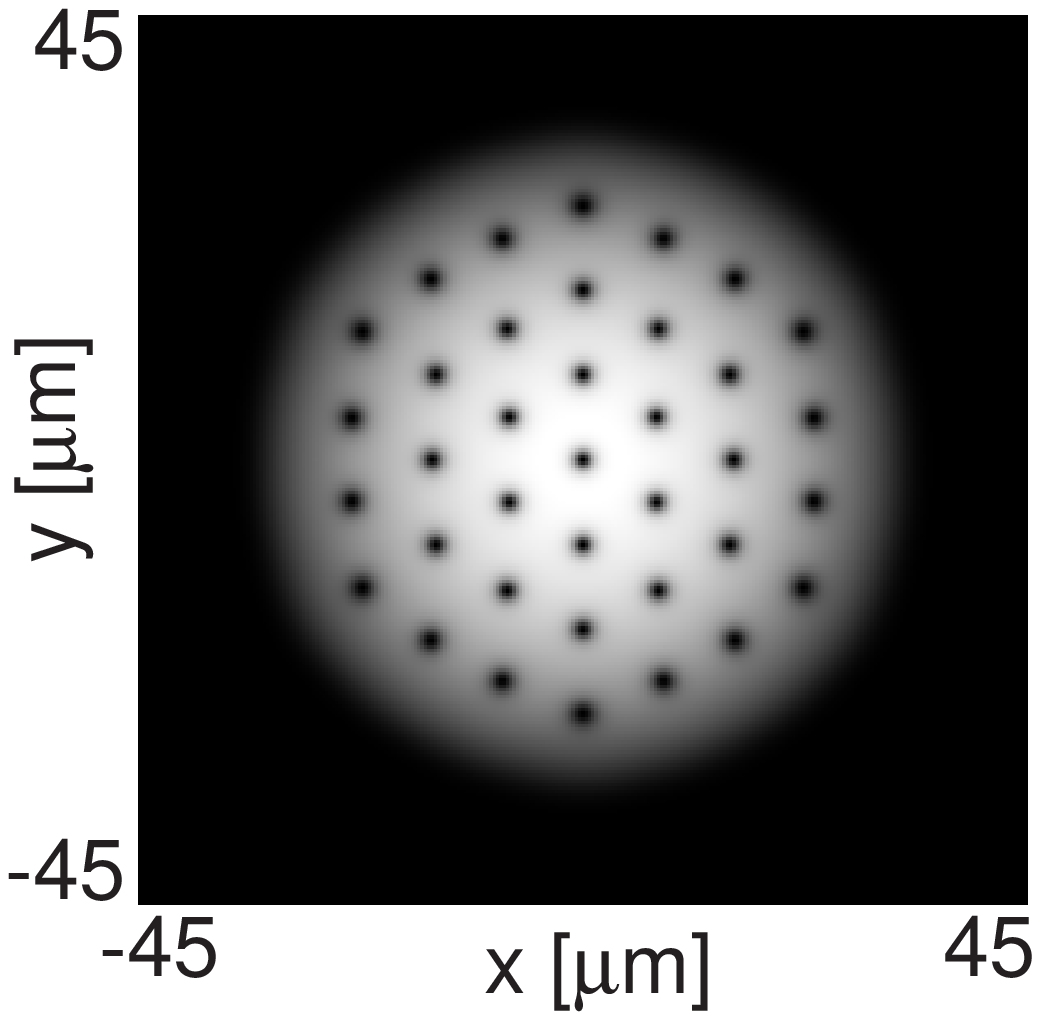} \\
(b) \includegraphics[width=6.5cm]{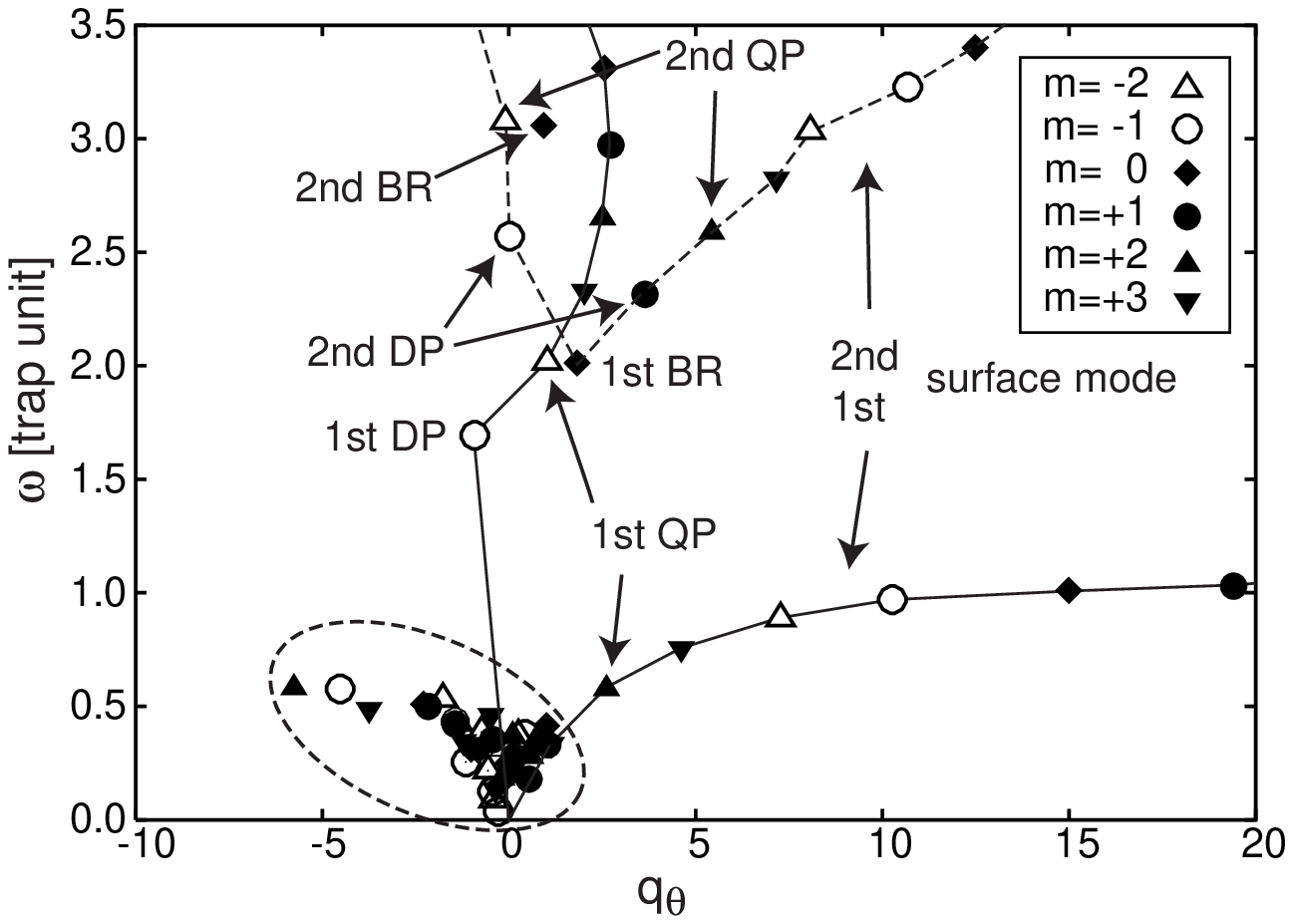} \\
(c) \includegraphics[width=6.5cm]{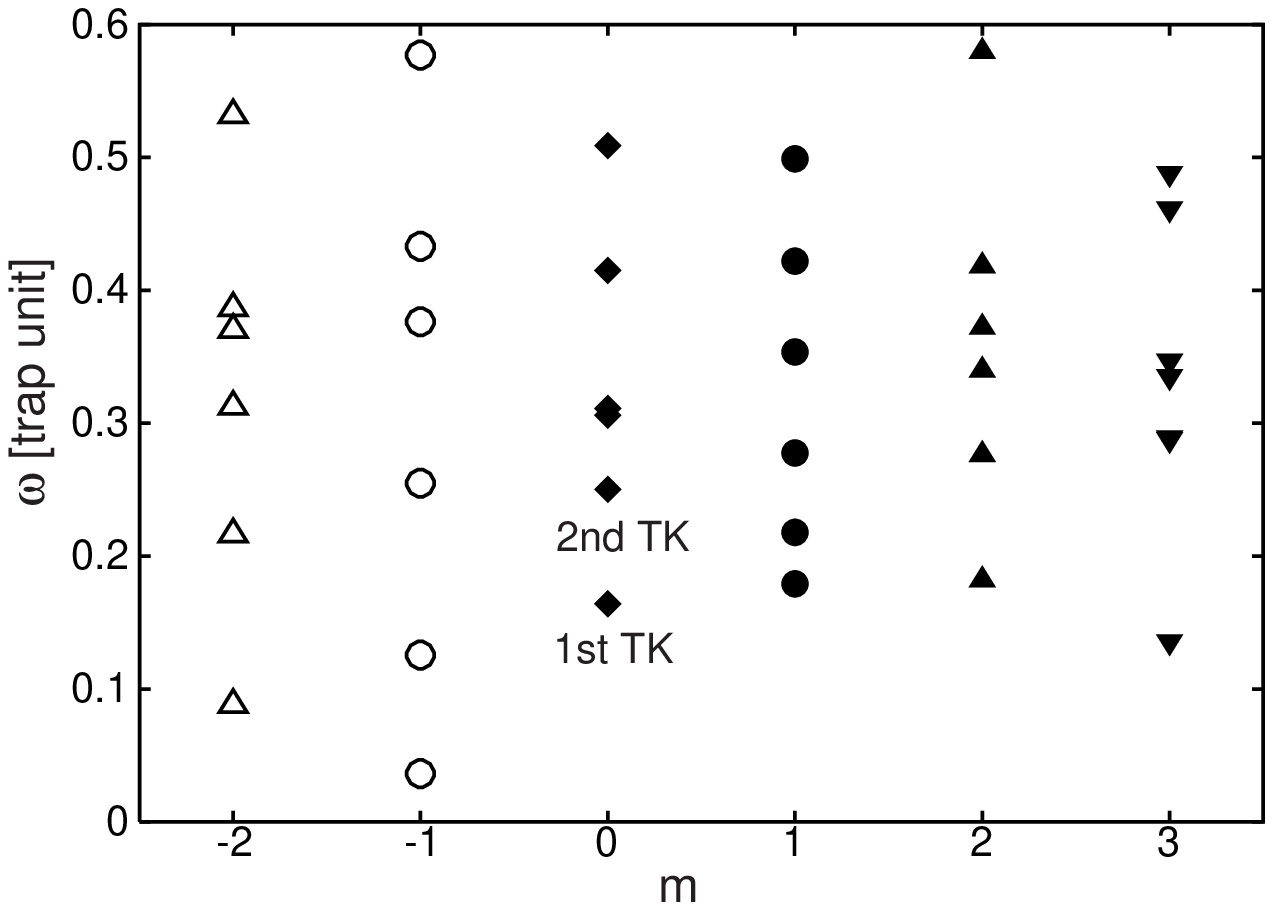} 
\caption{(a) Density profile of the condensate with 37 vortices at $\Omega = 0.7 \omega _r$.
(b) Collective excitations up to $\omega = 3.5\omega _r$.  
(c) Lowest excitations marked by a dotted line as functions of $q_{\theta}$
are displayed as functions of the symmetry index $m$.
Here, TK, BR, DP, and QP denote the Tkachenko, breathing, dipole, 
and quadrupole modes, respectively.
}
\label{fig:}
\end{figure}

The low-energy excitations are illustrated in Fig.~1. 
Here, the equilibrium states are found numerically via imaginary time propagation 
of Eq.~(\ref{eq:2dgp}); $t\rightarrow \tau = - it$. The initial vortex configuration 
is taken as a regular array with sixfold symmetry. The resulting configuration 
at $\Omega = 0.7\omega _r$ is formed by 37 vortices, displayed in Fig.~1(a).
In Fig.~1(b), the excitation energies up to $\omega=3.5\omega _r$ 
are shown as functions of the average angular momentum $q_{\theta}$.
Each collective mode is classified by a symmetry index, $m$, 
obtained from the above function, $F^{(u)}_{\bf q}(m)$, 
which characterizes the oscillation pattern.
For example, the breathing (BR), dipole (DP), and quadrupole (QP) modes have 
$m=0$, $\pm 1$, $\pm 2$, respectively.
The branch extending from the origin towards higher $q_{\theta}$-values 
consists of surface modes which are spaced periodically with $m$ (mod 6), 
where the outer condensate surface oscillates, 
the vortex cores being almost stationary at their equilibrium positions.
The other branch situated at higher energies and starting at $\omega = 2\omega _r$
represents another class of surface modes characterized by a node along the radial direction.
The energy spectra for the low-lying modes are depicted in Fig.~1(c) 
as functions of the symmetry index $m$. These low-lying eigenstates are the lattice-oscillation 
modes coupled with the condensate motion and lead to the distortion of both the lattice and 
the condensate surface towards an $m$-fold symmetric shape.
In addition to the lowest and second-lowest TK with $m=0$,  embedded among the other
modes, we found a parallel-precession mode with $m=-1$ and having the lowest energy 
where all the vortices precess in phase. 
For increasing energy, the modes for each $m$ feature nodes along $r$ and/or $\theta$.
The TK mode (with $m=0$) is selectively excited by the Gaussian laser beam with 
$0$-fold symmetry in the experiment \cite{coddington}. Likewise, a mode with $m=\pm \ell$
may be excited by a disturbance (e.g., magnetic) with $\ell$-fold symmetry \cite{memo}.

\begin{figure}[t]
\begin{tabular}{cc}
\includegraphics[width=4cm]{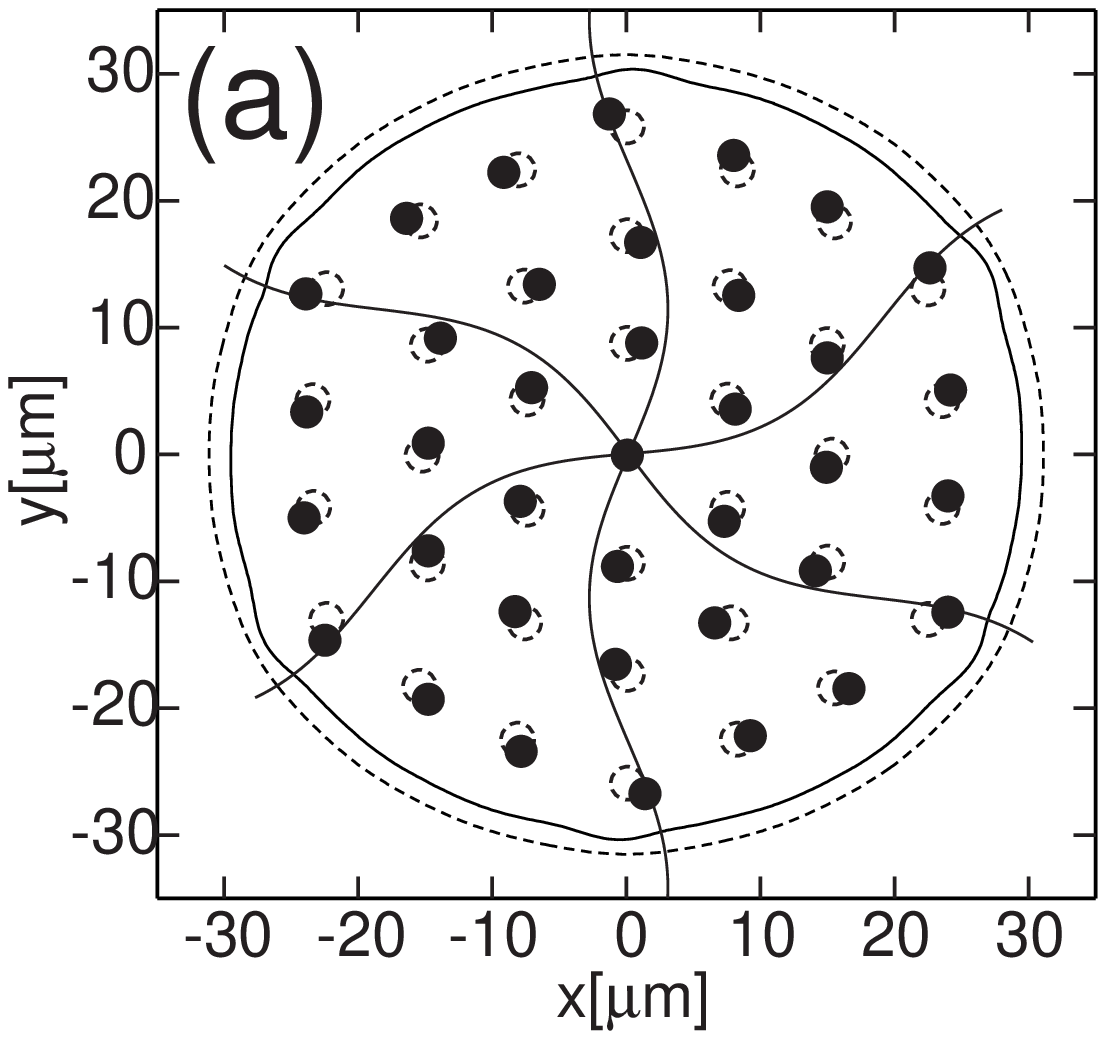} &
\includegraphics[width=4cm]{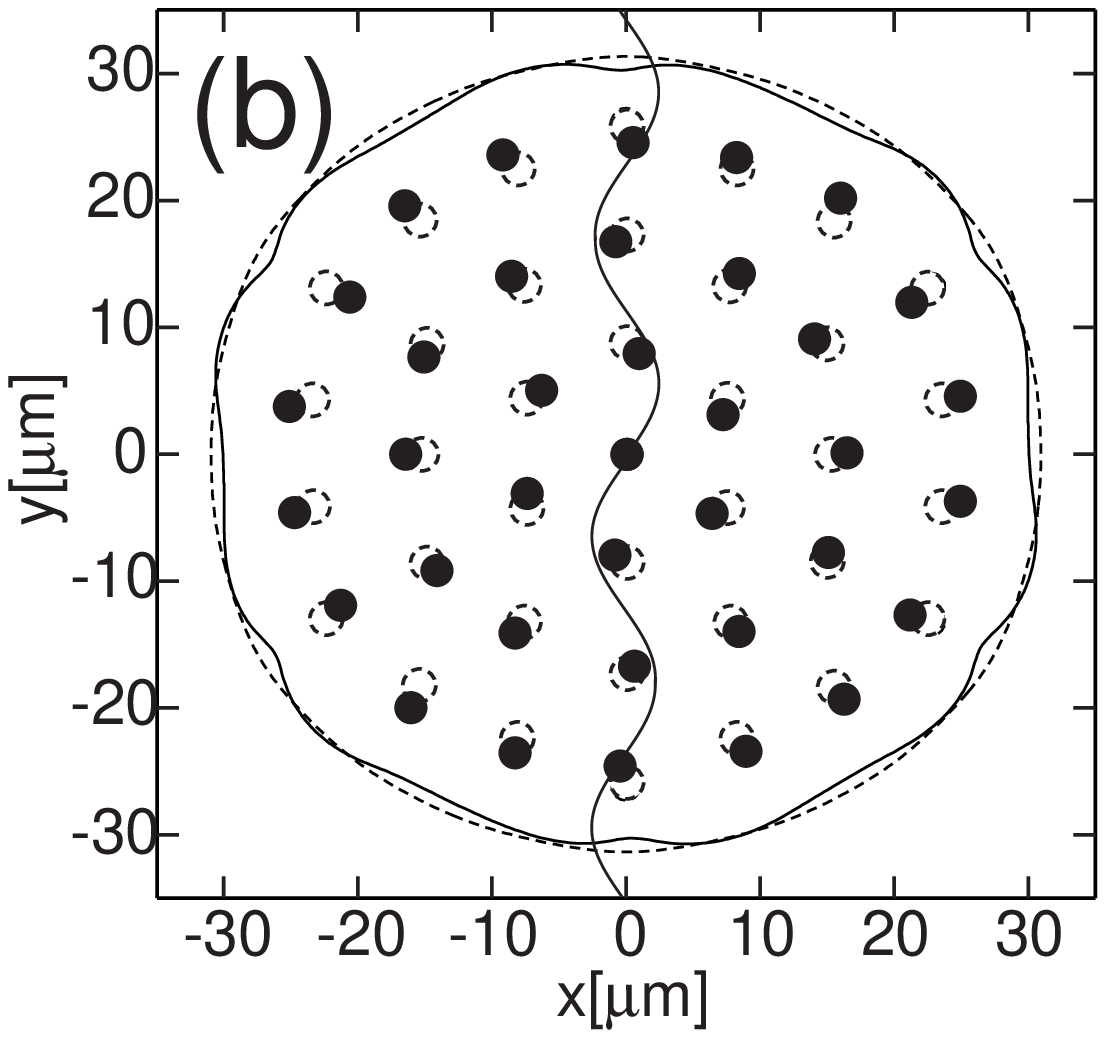} \\
\includegraphics[width=4cm]{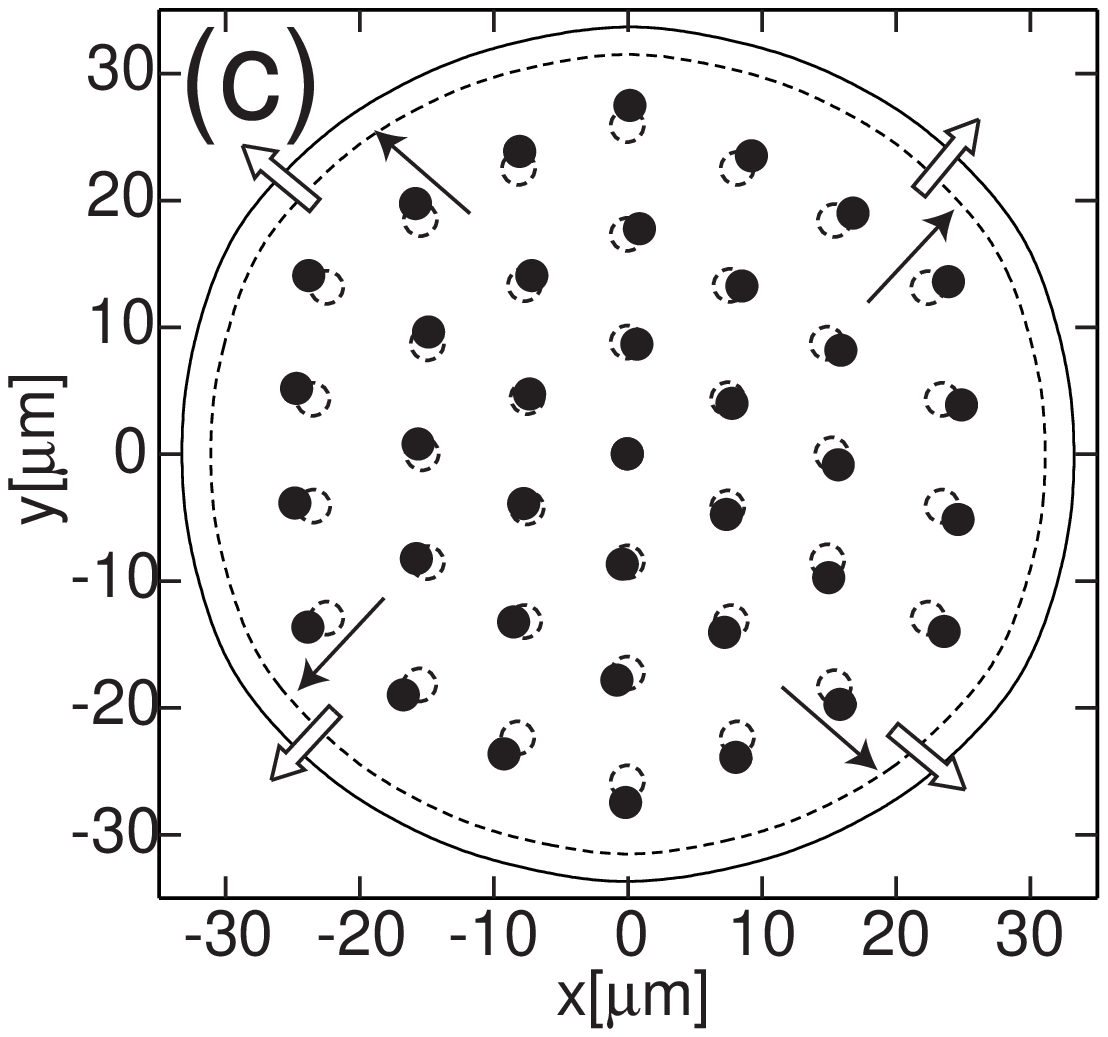} &
\includegraphics[width=4cm]{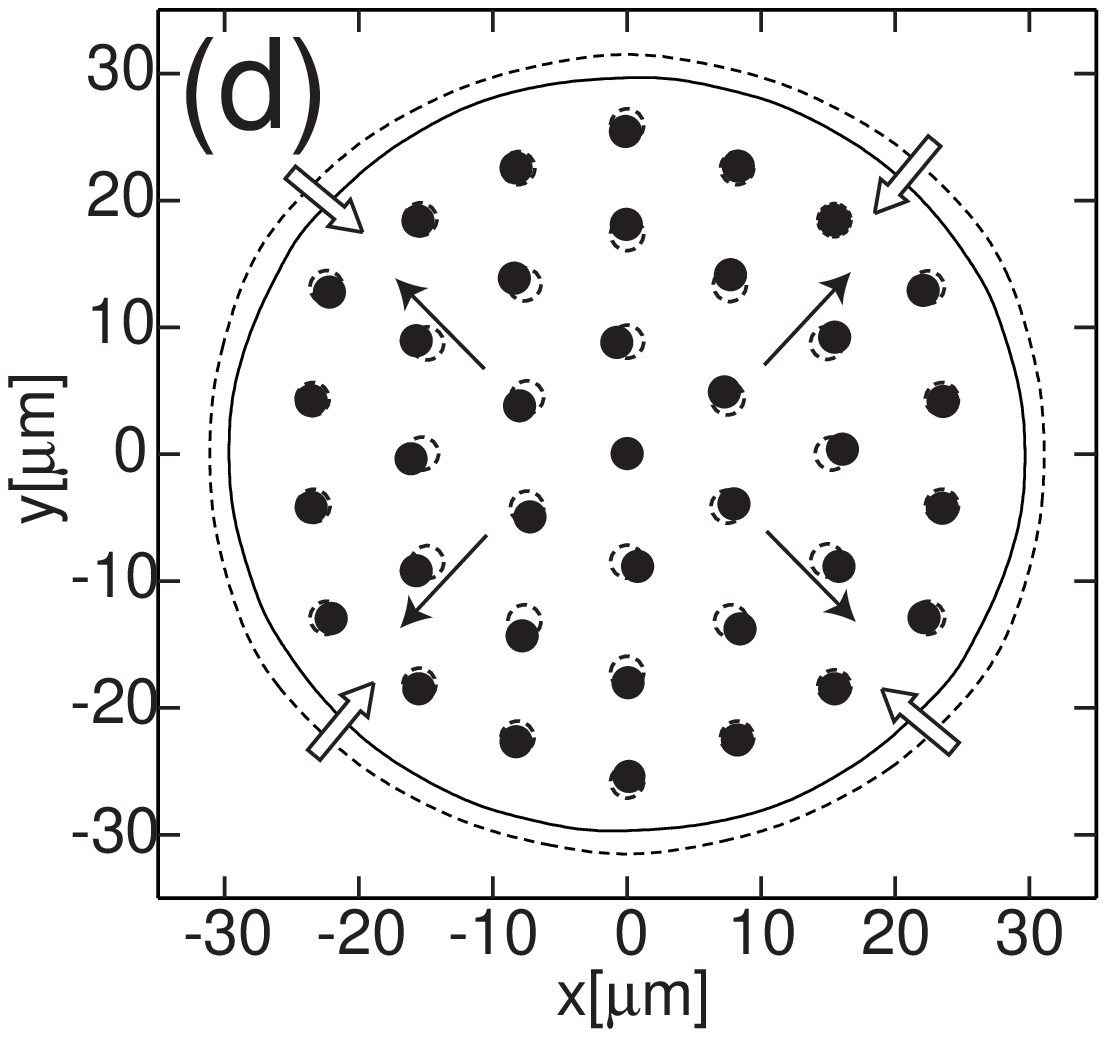} \\
\end{tabular}
\caption{Oscillation patterns for the (a) lowest TK, (b) second TK, 
(c) lowest BR, and (d) second BR modes are indicated by 
the filled circles and solid lines. Empty circles and dotted lines 
correspond to the equilibrium state, $\phi _g$.}
\label{fig:TK}
\end{figure}

Figure 2 describes the oscillation patterns for selected collective modes
obtained from a linear fluctuation $\delta \psi({\bf r},t)$: 
The lowest TK mode $\omega _{1,0}$ is shown in Fig.~2(a) where the empty (filled) circles
indicate the equilibrium (quarter-period) positions of the vortex cores.
The line is a fitted sinusoidal modulation wave (for clarity, the amplitude is magnified 
by a factor of $30$).
It is observed that the modulation displays a node at $r_{\rm node}\simeq 0.64R_{\rm TF}$,
which is in fair agreement with the JILA data of $r_{\rm node}=0.665 R_{\rm TF}$.
The second TK mode  $\omega _{2,0}$ is shown in Fig.~2(b).
It is seen that in contrast to the first TK mode whose core modulation is 
almost purely transversal, the second TK contains a longitudinal component, in addition to
the transverse one. Thus, the core executes elliptic motion around its equilibrium position.
This behavior coincides with the general trend for the ellipticity of a TK mode,
namely the expression for the amplitude ratio of the longitudinal and transverse components 
$\delta _{\rm L} / \delta _{\rm T} \propto 1/\sqrt{\Omega}k_{\rm TK}$,
where $k_{\rm TK}$ is the wavelength of the TK mode \cite{donnelly}. Therefore, 
the second TK with a shorter $k_{\rm TK}$ exhibits more pronounced longitudinal motion.
We also point out that via encountering the condensate radius in Figs.~2(a) and 2(b),
the TK modes tend to accompany the condensate motion.
We also observe that the ratio of the energies for these TK modes 
is $\omega _{2,0}/\omega _{1,0}= 1.56$ at $\Omega = 0.7\omega _r$, 
which compares favorably with the experimental data 
of $1.8\pm 0.2$ at $\Omega = 0.95\omega _r$ \cite{coddington} 
and with the value 1.63 obtained by Anglin and Crescimanno \cite{anglin},
who assert that it would be independent of $\Omega$.

As typical vortex-lattice oscillations apart from the TK, 
we now consider two contrasting compressional modes
which belong to the totally symmetric ($m=0$) ``breathing" branch.
Figures~2(c) and 2(d) depict the oscillation patterns for the lowest (1BR) 
and the second-lowest (2BR) breathing modes in Fig.~1(b).
The former (latter) BR mode executes mutual in-phase (out-of-phase) motion
between the cores and the condensate surface, as indicated with the two parallel (opposite) arrows
in the figure. The inner (outer) arrow denotes the motion of the cores (condensates).
Therefore, the former belongs to the common longitudinal mode and the latter
to the differential longitudinal mode.
It should be pointed out that the inner core motion in Figs.~2(c) and 2(d) is 
predominantly transversal; consequently, 
even in the nominally longitudinal BR mode all the vortex motion 
is not necessarily purely longitudinal.
This may be related to the `$s$-bending' \cite{coddington}
of the third observed mode.

Having obtained the complete spectrum of the low-lying excitations, 
let us now proceed to extend the analysis into nonlinear dynamics 
with real-time evolution simulated with the TDGP Eq.~(\ref{eq:2dgp}).
Since Coddington {\it et al.} excite the TK mode by shining resonant laser light 
in the center of the BEC to produce an inward flow, 
we introduce a Gaussian potential $V({\bf r})=-\hbar\omega _r e^{-r^2/(R_{\rm TF}/2)^2}$
localized in the center of the trap for a certain period 
and follow the time evolution according to Eq.~(1).

We illustrate the resulting oscillation patterns and their Fourier analyses in Fig.~3.
At $\Omega=0.7\omega _r$, the 37 vortices shown in Fig.~1(a) form 
a concentric regular lattice
around the central vortex, consisting of three circles, $j=1, 2, 3$.
We analyze the transverse vortex motion 
${\bf r}_{j,n}(t)=(x_{j,n}(t), y_{j,n}(t))$ in terms of the averaged angle 
$\theta _j (t) 
= \frac{1}{6}\sum^{5}_{n=0}[\arctan{(y_{j,n}(t)/x_{j,n}(t))} 
-\arctan{(y_{j,n}(0)/x_{j,n}(0))} ]$ 
where $n$ denotes the vortices aligned along the line and extending the angle $n\pi/3$
from the vertical, see Fig.~1(a). It is apparent in Fig.~3(a)
where we plot the time dependence of the vortex motion for each circle, $j=1, 2, 3$,
that (i) there exist several superposed oscillations. 
(ii) The outermost vortices ($j=3$) are in opposite phase with the inner vortices ($j=1, 2$).
This result coincides with the above calculations based on the BdG equations.
In fact, the nodal positions obtained from both calculations agree quite well.

\begin{figure}[t]
(a) \includegraphics[width=6.4cm]{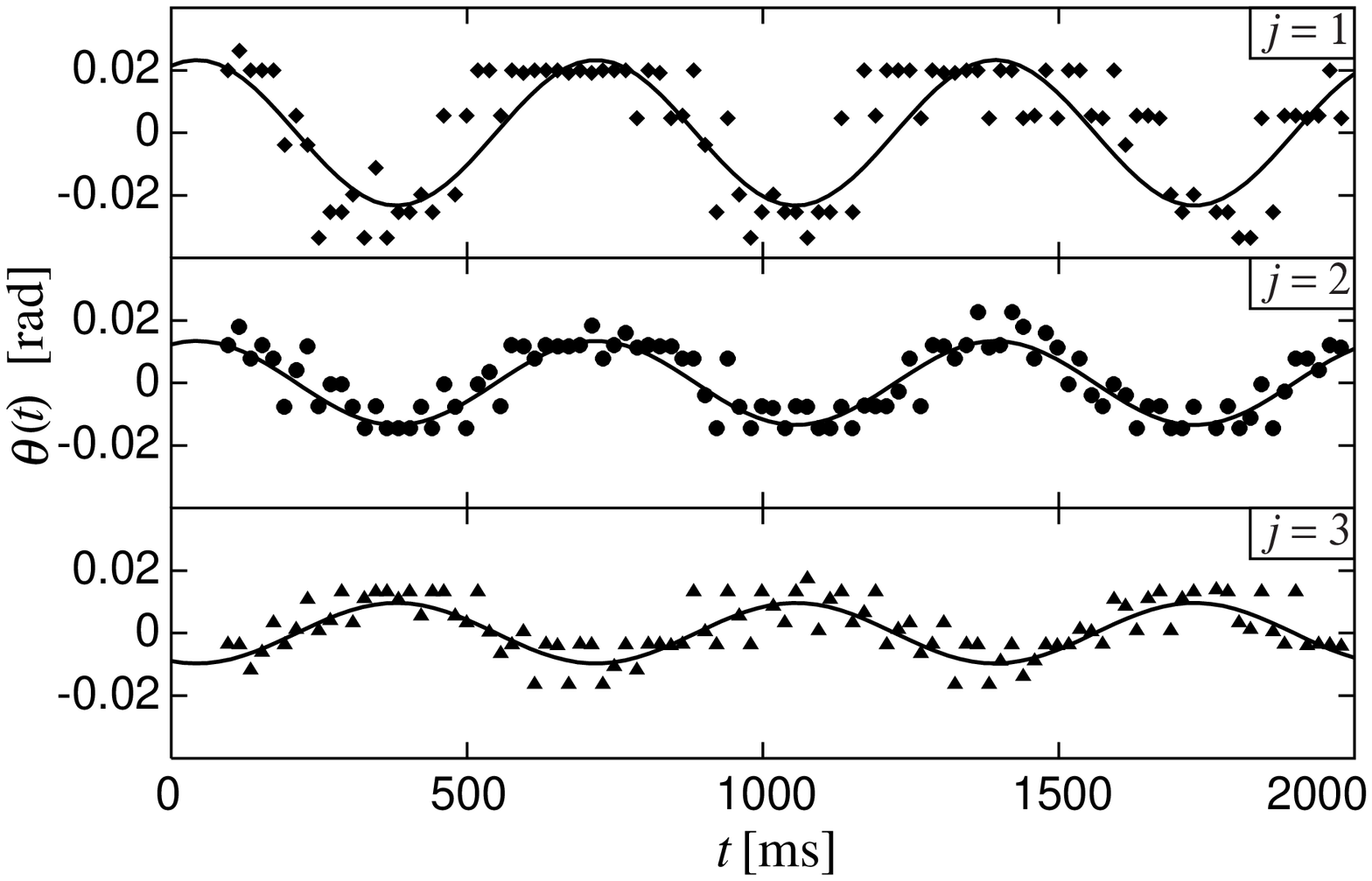} \\
(b) \includegraphics[width=6.4cm]{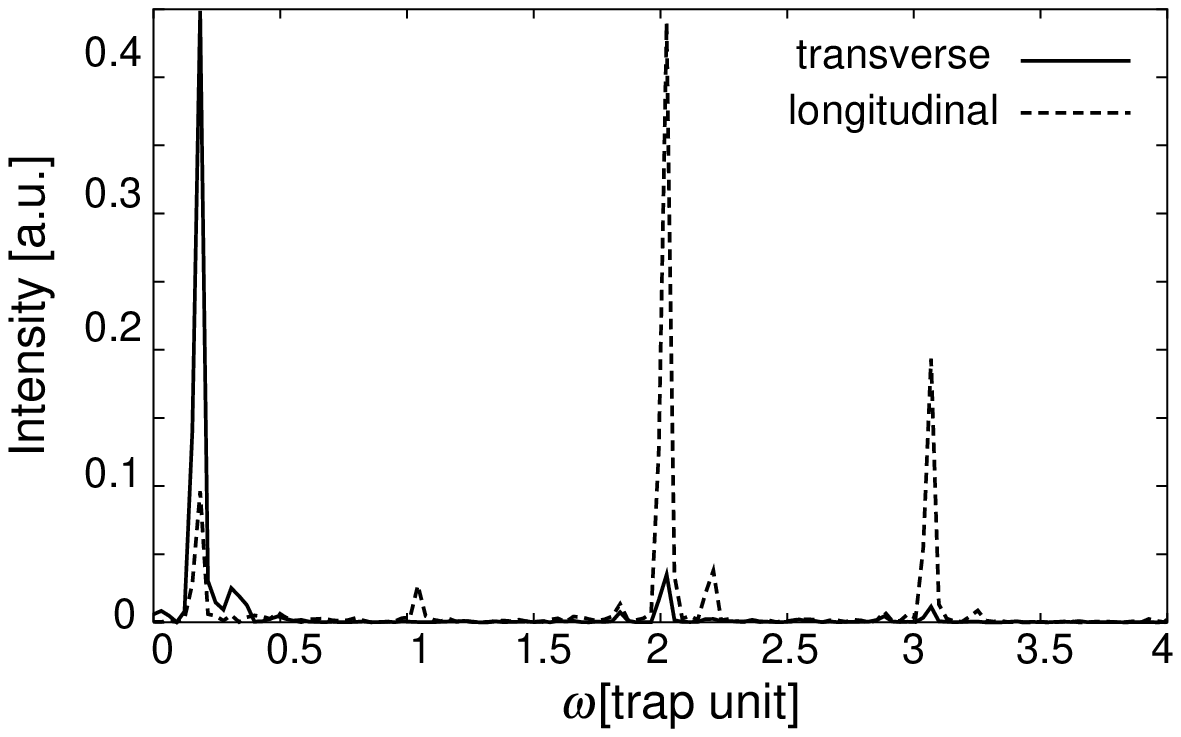} 
\caption{
(a) Transverse oscillations of vortices on three concentric circles, 
$j=1, 2, 3$, from the trap center.
(b) Fourier analyses of these oscillation patterns, $\theta _j (t)$, 
and the longitudinal vortex motion, $r_j(t)$.
}
\label{fig:}
\end{figure}

In Fig.~3(b), the Fourier analyses of these transverse oscillations, $\theta _j (t)$, and 
the longitudinal motions, $r_j (t)=\frac{1}{6}\sum^{5}_{n=0}|{\bf r}_{j,n}(t)|$, 
are depicted. It is seen
that the sharp peak at $\omega = 0.16\omega _r$ precisely coincides with  
the first TK mode $\omega_{1,0}$ identified above, see Fig.~1(c).
The second peak at $\omega = 2.0\omega _r$ corresponds to the 1BR mode,
belonging to the common longitudinal modes.
The third peak at $\omega \sim 3.1\omega _r$ may be identified as the 2BR mode,
see Figs.~1(b) and 2(d), and as a differential longitudinal mode.
These three collective modes closely match the observed characteristics.
In particular, the third mode which was tentatively assigned  
by Cozzini {\it et al.} and Choi {\it et al.} \cite{stringari}, independently, 
as a higher-order hydrodynamic mode, is now identified above.
It should be noted that the resonances of these three modes for a Gaussian potential 
have been numerically reproduced over a wide range of rotation rates $\Omega = 0.7 \sim 0.92$, 
corresponding to $37 \sim 121$ vortices.

In Fig.~4, we plot the first TK energy $\omega _{1,0}$ 
as a function of $\Omega/\omega _r$
together with the experimental data \cite{coddington} and a hydrodynamic
prediction by Anglin and Crescimanno \cite{anglin}.
There prevails close quantitative overall agreement 
between our results and the experimental data.  
We emphasize that our calculations contain no adjustable
parameters and also that our computations within the BdG and TDGP approaches 
agree within numerical accuracy $\sim 10^{-3}\omega _r$ 
below $\Omega \le 0.8\omega _r$.
Calculations for larger rotation rates, where BdG cannot be feasible from a numerical point
of view, are done with TDGP,
which enables us to extrapolate the BdG results to larger rotation rates.
The inset in Fig.~4 describes the $\Omega$ dependence of the 1BR and 2BR modes.
It is known that the former is consistent with an earlier prediction 
by Pitaevskii and Rosch \cite{pitaevskii},  who point out that the 2D BR mode 
features the universal eigenfrequency $\omega = 2.0\omega _r$.
Our result reproduces this result 
and further explains the observation mentioned above that the second 
peak in Fig.~3(b) is indeed the 1BR mode.
As for the third peak in Fig.~3(b), previously identified as the 2BR mode,
it is reckoned from this inset that the observed value $18.5\pm 0.3$Hz ($\simeq 2.2\omega _r$)
appears accountable, judging from the overall $\Omega$ dependence towards $\Omega _c$.

\begin{figure}[t]
\includegraphics[width=6.7cm]{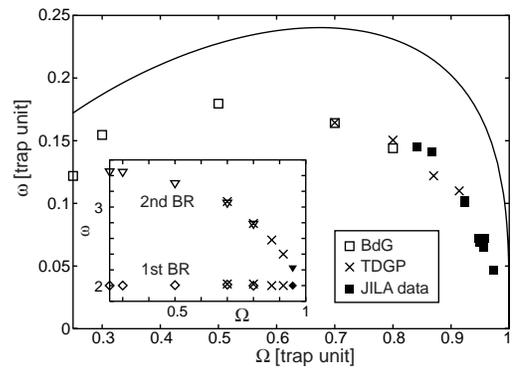}
\caption{
Frequencies of three modes obtained from the BdG and TDGP approaches 
are compared with the JILA data (filled squares, diamonds, and triangles) 
as functions of $\Omega$. In the main panel the TK (open squares), 
in the inset both the first BR (open diamonds) and the second BR (open triangles).
The solid line represents the dispersion relation in Ref. \cite{anglin}.}
\label{fig:}
\end{figure}

In summary, we have discussed the complete low-energy excitation spectrum in a vortex lattice
by solving the BdG equations. The $m=0$ subset of these solutions includes 
the transverse shear, common longitudinal, and differential longitudinal modes. 
We have also succeeded in simulating the actual experimental results, 
identifying a new pair of modes.

The authors thank V.\ Schweikhard, I.\ Coddington, and M.\ Ichioka
for useful conversations and communications.
KM is grateful to G.\ Baym, J.\ R.\ Anglin, S.\ Stringari, and A.\ L.\ Fetter
for enthusiastic discussions on the Tkachenko mode at the Aspen Center for Physics.

During the preparation of this manuscript, we learned about two closely related preprints 
\cite{baksmaty} and \cite{simula}. The former (latter) presents BdG (TDGP) 
treatments of the TK mode.


\begin{thebibliography}{99}

\bibitem{donnelly}
R.\ J.\ Donnelly, 
{\it Quantized Vortices in Helium II} 
(Cambridge University Press, Cambridge, 1991).

\bibitem{nist} 
M.\ R.\ Matthews {\it et al.}, Phys.\ Rev.\ Lett.\ {\bf 83}, 2498 (1999).

\bibitem{mit} 
A. E.\ Leanhardt {\it et al.},  Phys.\ Rev.\ Lett.\ {\bf 89}, 190403 (2002).

\bibitem{ens} 
K.\ W.\ Madison {\it et al.}, Phys.\ Rev.\ Lett.\ {\bf 84}, 806 (2000).

\bibitem{holland} 
J.\ E.\ Williams and M.\ J.\ Holland, Nature (London) {\bf 401}, 568 (1999).

\bibitem{nakahara}
M.\ Nakahara {\it et al.},  Physica B {\bf 284-288}, 17 (2000).

\bibitem{tkachenko}
V.\ K.\ Tkachenko, Sov.\ Phys.\ JETP {\bf 29},  245 (1969).

\bibitem{baym} 
G.\ Baym, Phys.\ Rev.\ B {\bf 51}, 11697 (1995).

\bibitem{fetter}
M.\ R.\ Williams and A.\ L.\ Fetter, Phys.\ Rev.\ B {\bf 16}, 4846 (1977). 

\bibitem{campbell}
L.\ J.\ Campbell, Phys.\ Rev.\ A {\bf 24}, 514 (1981).
  
\bibitem{sonin}
E.\ B.\ Sonin, Rev.\ Mod.\ Phys.\ {\bf 59}, 87 (1987).

\bibitem{andereck} 
C.\ D.\ Andereck and W.\ I.\ Glaberson, 
J.\ Low Temp.\ Phys.\ {\bf 48}, 257 (1982).
  
\bibitem{coddington}
I.\ Coddington {\it et al.}, Phys.\ Rev.\ Lett.\ {\bf 91}, 100402 (2003).

\bibitem{memo}
This has been confirmed by our numerical simulations of the TDGP equations 
for $\ell=0, 1, 2$.

\bibitem{anglin}
J.\ R.\ Anglin and M.\ Crescimanno, cond-mat/0210063.  

\bibitem{baym2}
G.\ Baym, Phys.\ Rev.\ Lett.\ {\bf 91}, 110402 (2003).

\bibitem{stringari}
M.\ Cozzini and S.\ Stringari, Phys.\ Rev.\ A {\bf 67}, 041602 (2003);
S.\ Choi {\it et al.}, Phys.\ Rev.\ A {\bf 68}, 031605 (2003).
  

\bibitem{leggett}
A.\ J.\ Leggett, Rev.\ Mod.\ Phys.\ {\bf 73}, 307 (2001).

\bibitem{isoshima}
T.\ Isoshima {\it et al.}, Phys.\ Rev.\ A {\bf 68}, 033611 (2003).

\bibitem{pitaevskii}
L.\ P.\ Pitaevskii and A.\ Rosch, Phys.\ Rev.\ A {\bf 55}, R853 (1997).

\bibitem{baksmaty}
L.\ O.\ Baksmaty {\it et al.}, cond-mat/0307368.

\bibitem{simula}
T.\ P.\ Simula {\it et al.}, cond-mat/0307130.

\end{thebibliography}
\end{document}